\documentstyle[aps,preprint,epsf]{revtex}
\newcommand{\beq}{\begin{equation}}
\newcommand{\eeq}{\end{equation}}
\begin{document}
\draft
\tightenlines

\title{ Investigation of the Field-Tuned Quantum Critical Point in
CeCoIn$_5$}

\author{ V.R. Shaginyan \footnote{E--mail:
vrshag@thd.pnpi.spb.ru}}
\address{$^a$Petersburg Nuclear Physics Institute,
Gatchina, 188300, Russia}
\maketitle

\begin{abstract}

The main properties and the type of the field-tuned quantum critical
point in the heavy-fermion metal CeCoIn$_5$ arisen upon applying
magnetic fields $B$ are considered within the scenario based on the
fermion condensation quantum phase transition. We analyze the behavior
of the effective mass, resistivity, specific heat, charge and heat
transport as functions of applied magnetic fields $B$ and show that in
the Landau Fermi liquid regime these quantities demonstrate the
critical behavior which is scaled by the critical behavior of the
effective mass.  We show that in the high-field non-Fermi liquid
regime, the effective mass exhibits very specific behavior, $M^*\sim
T^{-2/3}$, and the resistivity demonstrates the $T^{2/3}$ dependence.
Finally, at elevated temperatures, it changes to $M^*\sim T^{-1/2}$,
while the resistivity becomes linear in $T$.
In zero magnetic field,
the effective mass is controlled by temperature $T$, and the
resistivity is also linear in $T$.
The obtained results are
in good agreement with recent experimental facts.

\end{abstract}

\pacs{PACS numbers: 71.10.Hf, 71.27.+a, 74.72.-h}

Magnetic-field tuning of quantum critical points (QCPs) in heavy-fermion
(HF) metals becomes a subject of intense current interest because,
as it is widely accepted, an understating of quantum criticality
can clear up a mystery of fundamental physics of strongly
correlated systems \cite{ste}.
A fundamental question is whether the QCPs observed in HF metals are
different and related to different quantum phase transition or their
nature can be captured by the physics of a single quantum phase
transition. To answer this question, we have at least to explore a
particular quantum critical point in order to identify its nature. It
can hardly be done on pure theoretical grounds since there can exist a
great diversity of quantum phase transitions and corresponding QCPs in
nature \cite{sac,voj}. Therefore, mutually complementary experimental
facts related to the critical behavior and collected in measurements on
the same HF metal are of crucial importance for understanding
the physics of HF metals. Obviously, such HF metal is to exhibit
the critical behavior and has no additional phase transitions.
For example, the HF metal CeRu$_2$Si$_2$ can be regarded as fit for
such study because the measurements have shown neither evidence of
the magnetic ordering, superconductivity nor conventional Landau
Fermi-liquid (LFL) behavior down to ultralow temperatures \cite{tak}.
Unfortunately, by now only precise ac susceptibility and static
magnetization measurements at small magnetic fields and ultralow
temperatures are known \cite{tak}. While additional measurements of such
properties as the heat and charge transport and  the specific
heat could produce valuable information about the existence
of Landau quasiparticles and their degradation and clarify the role
of the critical fluctuations near the corresponding QCP. Such
measurements on the HF metal CeCoIn$_5$ were recently reported
\cite{pet,pag,bi,pag1}. It was shown that the resistivity $\rho(T)$
of CeCoIn$_5$ as a function of temperature $T$ is linear in $T$
in the absence of magnetic field \cite{pet}. Due to the existence of
magnetic field-tuned QCP with a critical field $B_{c0}\simeq 5.1$ T,
the LFL behavior is restored at magnetic fields $B\geq B_{c0}$
\cite{pag,bi,pag1}. At the LFL regime, the measurements of
the specific heat and the coefficient $A$ in the resistivity,
$\rho(T)=\rho_0+A(B)T^2$, describing the electron-electron scattering,
have demonstrated  that the Kadowaki-Woods ratio,
$K=A(B)/\gamma^2(B)$ \cite{kadw}, is conserved \cite{bi}.
Here $\gamma(B)=C/T$, and $C$ is the specific heat. It was also shown
that the coefficient $A$ diverges as $A(B)\propto
(B-B_{c0})^{-\alpha}$, with $\alpha\simeq 4/3$ \cite{pag,pag1}.
Moreover, a recent study of CeCoIn$_5$
in magnetic fields $B>B_{c0}$ have
revealed that the coefficients $A(B)$ and $C(B)$,
with $C(B)$ describing a $T^2$ contribution to thermal resistivity
$\kappa_r$, possess the same critical field dependence $A(B)\propto
C(B)\propto{(B-B_{c0})^{-\alpha}}$, so that
the ratio $A(B)/C(B)=c$  \cite{pag1}. Here $c$ is a field-independent
constant characterizing electron-electron scattering in metals and
having a typical value of $0.47$, see e.g. \cite{ben,pag2}.
The same study has discovered that the resistivity
behaves as $\rho(T)\propto T^{n}$ in the high-field non-Fermi
liquid (NFL) regime, with $n\simeq 2/3$, while in the
low-field NFL regime, at $B\sim B_{c0}$, the exponent $n\simeq 0.45$
\cite{pag1}. Note that the same behavior of the resistivity
was observed in the HF metals URu$_2$Si$_2$ \cite{kim} and YbAgGe
\cite{bud} on the verge of the LFL regime, and
that the critical behavior
takes place up to rather high temperatures comparable with the
effective Fermi temperature $T_k$ and up to high magnetic fields.
For example, the resistivity measured on
CeCoIn$_5$ shows the $T^{2/3}$ behavior over one
decade in temperature from 2.3 K to 20 K,
and the coefficients $A(B)$ and $C(B)$
exhibit the same behavior  at the fields
from $B=B_{c0}=5.1$ T to at least 16 T  \cite{pag1}.

In this Letter, we present an explanation of the observed
behavior of the electronic system of
the heavy-fermion metal CeCoIn$_5$ arisen upon applying
magnetic fields $B$. We analyze the behavior
of the effective mass, resistivity, specific heat, charge and heat
transport as functions of the applied magnetic field $B$
and show that in the Landau Fermi liquid regime, these quantities
demonstrate the critical behavior which is scaled by the critical
behavior of the effective mass.
In that case, the critical behavior is determined by the fermion
condensation quantum phase transition (FCQPT), whose physics is
controlled by quasiparticles with the effective mass which strongly
depends on the applied magnetic field $B$ and diverges at $B\to
B_{c0}$. In zero magnetic field,
the effective mass is controlled by temperature $T$, and the
resistivity is linear in $T$.
In the high-field non-Fermi liquid regime
when the system
comes from the LFL behavior to the NFL one, the effective mass exhibits
very specific behavior, $M^*\sim T^{-2/3}$, and the resistivity
demonstrates the $T^{2/3}$ dependence. In the low-field NFL regime, at
$B\sim B_{c0}$, this behavior becomes complicated so that the
resistivity behaves as $T^{n}$, with $n\sim 0.7-0.8$.
At elevated temperatures and in zero magnetic field, the behavior
changes to $M^*\sim T^{-1/2}$, while the resistivity becomes linear in
$T$.

We start with a brief consideration of
the LFL regime restored by the application of magnetic field
$B>B_{c0}$. If the electronic system approaches FCQPT from the
disordered side, the effective mass $M^*(B)$ of the restored LFL
depends on magnetic field $B$ as \cite{shag1,zkh}
\beq M^*(B)\propto
\frac{1}{(B-B_{c0})^{2/3}}.\eeq
Note that Eq. (1) is valid at $T\ll T^*(B)$, where
the function $T^*(B)\propto
(B-B_{c0})^{4/3}$ determines the line on the $B-T$ phase diagram
separating the region of the LFL behavior from the NFL behavior taking
place at $T>T^*(B)$ \cite{shag1}. To estimate the coefficient $A$, we
observe that at the highly correlated regime when $M^*/M\gg 1$, the
coefficient $A\propto (M^*)^2$, here $M$ is the bare electron mass
\cite{khsc}.  As a result, we have
\beq A^2(B)\propto
\frac{1}{(B-B_{c0})^{4/3}},\eeq
and observe that in the LFL regime, the Kadowaki-Woods ratio,
$K=A(B)/\gamma^2(B)$, is conserved because $\gamma(B)\propto M^*(B)$.

Let us now turn to consideration of the system's behavior
at elevated temperatures paying special attention to the transition
region. To do it, we use the well-know Landau equation relating
the quasiparticle energy $\varepsilon({\bf p})$ near the
Fermi surface to variations $\delta n({\bf p},T)$ of the
quasiparticle distribution function $n_F({\bf p},T)$ \cite{lan,lanl1}
\beq \varepsilon({\bf p})-\mu=\frac{p_F(p-p_F)}{M^*}+
\int F({\bf p},{\bf p}_1)\delta n({\bf p}_1,T)
\frac{d{\bf p}_1}{(2\pi)^3}.\eeq
Here, $\mu$ is the chemical potential, $p_F$ is the Fermi
momentum, $F({\bf p},{\bf p}_1)$ is the Landau amplitude.
For the
sake of simplicity the summation over the spin variables is omitted.
In our case, the variation $\delta n({\bf p},T)$ is induced by
temperature $T$ and defined as
$\delta n({\bf p},T)=n_F({\bf p},T)-n_F({\bf p},T=0)$ with
$n_F({\bf p},T)$ being given by the
Fermi-Dirac function \begin{equation} n_F({\bf p},T)=\left\{ 1+\exp
\left[ \frac{(\varepsilon({\bf p})-\mu )}T \right] \right\}^{-1}.
\end{equation}
Taking into account that
$\varepsilon(p\simeq p_F)-\mu=p_F(p-p_F)/M^{*}$,
one directly obtains from Eqs. (4) that $n_F({\bf p},T\to 0)\to \theta
(p_F-p)$, where $\theta (p_F-p)$ is the step function. In our case,
Eq. (3) can be used to estimate the behavior of the effective mass
$M^*(T)$ as a function of temperature. Actually, differentiating both
parts of Eq. (3) with respect to the momentum $p$, we observe that
the difference $p_F/M^*(T)-p_F/M^*(T=0)$ is given by the integral. In
its turn, the integral $I$ can be estimated upon using the standard
procedure of calculating integral when the integrand contains the
Fermi-Dirac function, see e.g.  \cite{lanl2}.  As a result, we obtain
that \beq \frac{M}{M^*(T)}\simeq
\frac{M}{M^*}+a_1\left(\frac{TM^*(T)}{T_kM}\right)^2
+a_2\left(\frac{TM^*(T)}{T_kM}\right)^4+...\eeq
Here $a_1$ and $a_2$ are constants proportional to the derivatives
of the Landau amplitude with respect to the momentum $p$.
Equation (5) can be regarded as
a typical equation of the LFL theory with the only exception
for the effective mass $M^*$ which strongly depends on the magnetic
field and diverges at $B\to B_{c0}$ as it follows from Eq. (1).
Nonetheless, at $T\to 0$, the corrections to $M^*(B)$ start with
$T^2$ terms provided that
\beq M/M^*(B)\gg a_1\left(\frac{TM^*(B)}{T_kM}\right)^2, \eeq
and the system exhibits the LFL behavior. At some temperature
$T_1^*(B)\ll T_k$, the value of the sum on the right hand side of Eq.
(5) is determined by the second term.
Then Eq. (6) is not valid, and upon omitting the first
and third terms,  Eq. (5) can be used to determine the effective mass
$M^*(T)$ in the transition region, \beq M^*(T)\propto T^{-2/3}.\eeq
We note, that Eq. (7) has been derived in \cite{zkh}.
Upon comparing Eq. (1) and Eq. (7) and taking into account that
the effective mass $M^*(T)$ is a continuous function of $T$, we can
conclude that $T_1^*(B)\propto (B-B_{c0})$.

A few remarks are in order here.
Equation (7) is valid if
the second term in Eq. (5) is much bigger then the first one, that is
\beq \frac{T}{T_k} \gg \left(\frac{M}{M^*}\right)^{3/2},\eeq
and this term is bigger then the third one,
\beq \frac{T}{T_k} \ll \frac{M}{M^*}.\eeq
Obviously, both Eq. (8) and (9) can be simultaneously satisfied
if $M/M^*\ll 1$. It is
seen from Eqs. (1) and (9) that at $B\to B_{c0}$, the range of
temperatures over which Eq. (7) is valid shrinks to zero,
as well as $T_1^*(B)\to 0$.
Thus, it is possible to observe the behavior
of the effective mass given by Eq. (7) in a wide range of temperatures
provided that the effective mass $M^*(B)$ is diminished by the
application of the high magnetic field, see Eq. (1).
At $B\to B_{c0}$ and finite
temperatures, Eq. (9) cannot be satisfied. Therefore, at elevated
temperatures, the third term comes into play making the function
$M^*(T)$ be complicated.  To estimate the exponent $n$, we take into
account only the third term in Eq.  (5) and obtain $M^*(T)\propto
T^{-n}$, with $n=4/5$. As a result, at $B\to B_{c0}$ and
$T>T_1^*(B)$, we have an approximation \beq
M^*(T)\propto T^{-n},\eeq with the exponent $n\sim 0.7-0.8$. The
contribution coming from the other terms can only enlarge the exponent.
On the other hand, $n<1$ because behind FCQPT, when the fermion
condensate is formed, $M^*(T)\propto 1/T$ \cite{duck}.
Detailed analysis of this item will be published elsewhere.
Then, at elevated temperatures, the system comes to a different regime.
Smoothing out the step function $\theta(p_F-p)$ at
$p_F$, the temperature creates the variation $\delta n({\bf p})\sim 1$
over the narrow region $\delta p\sim M^*T/p_F$. In fact, the series on
the right hand side of Eq. (5) representing the value of the integral
$I$ in Eq. (3) is valid, provided that the interaction radius $q_0$ in
the momentum space of the Landau amplitude $F$ is much larger than
$\delta p$, $q_0\gg\delta p$.  Otherwise, if $q_0\sim\delta p$, the
series do not represent $I$ and Eqs. (5) and (7) are no longer valid.
Such a situation takes place at rising temperatures because the product
$M^*T$ grows up as $q_0\sim \delta p\sim M^*T/p_F\propto T^{1/3}$, as
it follows from Eq.  (7). As a result, the integral runs over the
region $q_0$ and becomes proportional to $M^*T/p_F$. Upon omitting the
first term on the right hand side of Eq. (3) and substituting the
integral by this estimation, we obtain the equation which determines
the behavior of the effective mass at $T>T^*(B)$ as \cite{shag1,shag2}
\beq M^*(T)\propto T^{-1/2}.\eeq

To capture and summarize the salient features of the LFL behavior
observed recently in CeCoIn$_5$ \cite{bi,pag1}, we apply
the above consideration based on FCQPT. The study of CeCoIn$_5$
in the LFL regime have
shown that the coefficients $A(B)$ and $C(B)$,
determining the $T^2$ contributions to the resistivity $\rho$
and thermal resistivity $\kappa_r$ respectively, possess the same
critical field dependence  \cite{pag1}
\beq A(B)\propto C(B)\propto\frac{1}{(B-B_{c0})^{4/3}}. \eeq
The observed critical exponent $4/3$ is in excellent
agreement with that of given by Eq. (2).
Such the parallel behavior of charge and heat transport
with the scattering rate growing as $T^2$
shows that the delocalized fermionic excitations are
the Landau quasiparticles carrying charge $e$.
We note that these should be destroyed in the case of
conventional quantum phase transitions
\cite{sac,voj}. Nonetheless, let us assume for a moment that
these survive. Since the heat and charge transport tend to
strongly differ in the presence of the critical fluctuations of
superconducting nature, the constancy of the ratio rules out the
critical fluctuations  \cite{pag1}. Therefore, we are led to the
conclusion that the observed value of the critical magnetic field
$B_{c0}=5.1$ T that coincides approximately with $H_{c2}=5$ T, the
critical field at which the superconductivity vanishes, cannot be
considered as giving grounds for the existence of quantum critical
behavior of new type.  Then, one could expect
that some kind of critical fluctuations could cause the observed
parallel behavior of charge and heat transport.
For example, it is impossible in the case of ferromagnetic fluctuations
with a wavevector $q\simeq 0$, but
large-$q$ scattering from antiferromagnetic fluctuations of finite
momenta could degrade the heat and charge transport in a similar way
\cite{pag2}. In this case, in order to preserve the Kadowaki-Woods
ratio these fluctuations are to properly influence the specific heat
which characterizes the thermodynamic properties of the system and is
not directly related to the transport one. On the other hand, there are
no theoretical grounds for the conservation of the Kadowaki-Woods ratio
within the frameworks of conventional quantum phase transitions
\cite{shag3}. Therefore, the conservation of the Kadowaki-Woods ratio
observed in recent measurements on CeCoIn$_5$ \cite{bi} definitely
seems to rule out these fluctuations.  While both the constancy of
Kadowaki-Woods ratio \cite{bi} and the constancy of the $A(B)/C(B)$
ratio \cite{pag1} give strong evidence in favor of the quasiparticle
picture.

Now we turn to consideration of the resistivity $\rho(T)$.
As we will see below, the striking recent measurements
of the resistivity \cite{pag1,kim,bud} furnish new evidence in favor of
the quasiparticle picture and the existence of FCQPT.

As it follows from Eq. (11) and the mention above relation $A\propto
(M^*)^2$, the term $AT^2\propto M^*T^2$ turns out to be $\propto T$
\cite{shag1}.  As a result, in zero magnetic field and relatively high
temperatures $T>T_c$, the resistivity of CeCoIn$_5$ is linear in $T$.
Here $T_c$ is the critical temperature at which the superconductivity
vanishes.
This observation is in good agreement with experimental facts
\cite{pet}.

At temperatures $T<T_1^*(B)$ and magnetic field $B>B_{c0}$, the system
exhibits the LFL behavior with the $T^2$ dependence of the
resistivity $\rho(T)$. Such a behavior is in agreement with
experimental facts \cite{pag,bi,pag1}.

At the high applied magnetic field and finite temperatures
$T>T_1^*(B)$ when the system comes into the NFL regime, the effective
mass $M^*$ is determined by Eq. (7). In that case, the range of
temperatures over which Eq. (7) is held  becomes rather wide, and the
system demonstrates the anomalous $T^{2/3}$ resistivity.
Actually, upon using the same arguments, we obtain that $AT^2\propto
(M^*)^2T^2\propto T^{2/3}$ and conclude that the resistivity
$\rho(T)\propto T^{2/3}$.  Again, this result is in excellent agreement
with the reported observations \cite{pag1,kim,bud}.

If the magnetic field $B\to B_{c0}$ and the temperature is
relatively high, $T>T_1^*(B)$, so that the system enters the NFL regime,
the effective mass is given by Eq. (10). In that case, the resistivity
$\rho(T)\propto (M^*)^2T^2\propto T^{k}$, with $k=2-2n=0.6-0.4$.
This result is in reasonable agreement with the reported observation
of anomalous $T^{0.45}$ dependence of the resistivity in a small region
near the critical field $B_{c0}=5.1$ T \cite{pag1}.

In conclusion, we have shown that
the experimentally observed behavior of the electronic system of
the heavy-fermion metal CeCoIn$_5$ arisen upon applying
magnetic fields can be understood within the frameworks of
the FCQPT scenario. We have shown that in the LFL
regime the resistivity, specific heat, charge and heat transport as
functions of the applied magnetic field $B$
demonstrate the critical behavior which is scaled by the critical
behavior of the effective mass.
We have observed that this critical behavior is determined by
FCQPT, whose physics is
controlled by quasiparticles with the effective mass which in
the LFL regime strongly depends on the applied magnetic field and
diverges at $B\to B_{c0}$. In zero magnetic field,
the effective mass is controlled by temperature $T$, and the
resistivity is linear in $T$. In the high-field NFL regime, the
effective mass exhibits very specific behavior, $M^*\sim T^{-2/3}$,
while the resistivity demonstrates the $T^{2/3}$ dependence. At
elevated temperatures, the behavior changes to $M^*\sim T^{-1/2}$,
while the resistivity becomes linear in $T$.


\begin{thebibliography}{99}

\bibitem{ste} G.R. Stewart, Rev. Mod. Phys. {\bf 73}, 797 (2001).

\bibitem{sac} S. Sachdev, {\it Quantum Phase transitions},
Cambridge, Cambridge University Press, 1999.

\bibitem{voj} M. Vojta, Rep. Prog. Phys. {\bf 66}, 2069 (2003).

\bibitem{tak} D. Takahashi {\it et al.,} Phys. Rev. B {\bf 67},
180407 (2003).

\bibitem{pet} C. Petrovic {\it et al.,} J. Phys. Condens. Matter
{\bf 13}, L337 (2001).

\bibitem{pag} J. Paglione {\it et al.,} Phys. Rev. Lett.
{\bf 91}, 246405 (2003).

\bibitem{bi} A. Bianchi {\it et al.,}
Phys. Rev. Lett. {\bf 91},  257001 (2003).

\bibitem{pag1} J. Paglione {\it et al.,} cond-mat/0405157.

\bibitem{kadw} K. Kadowaki and S.B. Woods, Solid State Commun.
{\bf 58}, 507 (1986).

\bibitem{ben} A.J. Bennet and M.J. Rice, Phys. Rev. {\bf 185}, 968
(1969).

\bibitem{pag2} J. Paglione, cond-mat/0404269.

\bibitem{kim} K.H. Kim {\it et al.,}
Phys. Rev. Lett. {\bf 91},  256401 (2003).

\bibitem{bud} S.L. Bud'ko, E. Morosan, and P.C. Canfield,
Phys. Rev. B {\bf 69}, 014415 (2004).

\bibitem{shag1}  V.R. Shaginyan, JETP Lett. {\bf 77}, 99 (2003); V.R.
Shaginyan, JETP Lett. {\bf 77}, 178 (2003).

\bibitem{zkh} M.V. Zverev and V.A. Khodel,
JETP Lett. {\bf 79}, 772 (2004).

\bibitem{khsc} V.A. Khodel and P. Schuck,
Z. Phys. B {\bf 104}, 505 (1997).

\bibitem{lan} L. D. Landau, Sov. Phys. JETP  {\bf 3}, 920 (1956).

\bibitem{lanl1}  E.M. Lifshitz and L.P. Pitaevskii,
{\it Statistical Physics,}
Part 2, Butterworth-Heinemann, 1999.

\bibitem{lanl2}  E.M. Lifshitz and L.P. Pitaevskii,
{\it Statistical Physics,}
Part 1, Butterworth-Heinemann, 2000, p. 168.

\bibitem{duck} J. Duckelsky {\it et al.,}
Z. Phys. B {\bf 102}, 245 (1997).

\bibitem{shag2}  V.R. Shaginyan, JETP Lett. {\bf 79}, 344 (2004).

\bibitem{shag3}  V.R. Shaginyan, J.G. Han, and J. Lee,
Phys. Lett. A, in press; cond-mat/0405025.

\end{thebibliography}
\end{document}